\documentclass[
reprint,
superscriptaddress,
showpacs,
 amsmath,amssymb,
floatfix,
]{revtex4-1}
\usepackage{dcolumn}
\usepackage{bm}
\usepackage{turnstile}

\usepackage{graphicx}
\usepackage{amsmath,amssymb}
\usepackage{url}

\usepackage{epsfig}
\usepackage{epstopdf}

\begin{document}

\title{
Magnetic proximity effect in [Nb/Gd] superlattices seen by neutron scattering
}

\author{Yu.~N.~Khaydukov}
\affiliation{Max-Planck-Institut f\"ur Festk\"orperforschung, Heisenbergstra\ss e 1, D-70569 Stuttgart, Germany}
\affiliation{Max Planck Society Outstation at the Heinz Maier-Leibnitz Zentrum (MLZ), D-85748 Garching, Germany}
\affiliation{Skobeltsyn Institute of Nuclear Physics, Moscow State University, Moscow 119991, Russia}

\author{E.~A.~Kravtsov}
\affiliation{Institute of Metal Physics, 620180 Ekaterinburg, Russia}
\affiliation{Ural Federal University, 620002 Ekaterinburg, Russia}

\author{V.~D.~Zhaketov}
\affiliation{Joint Institute for Nuclear Research, 141980 Dubna, Russia}

\author{V.~V.~Progliado}
\affiliation{Institute of Metal Physics, 620180 Ekaterinburg, Russia}

\author{G.~Kim}
\affiliation{Max-Planck-Institut f\"ur Festk\"orperforschung, Heisenbergstra\ss e 1, D-70569 Stuttgart, Germany}

\author{Yu.~V.~Nikitenko}
\affiliation{Joint Institute for Nuclear Research, 141980 Dubna, Russia}

\author{T.~Keller}
\affiliation{Max-Planck-Institut f\"ur Festk\"orperforschung, Heisenbergstra\ss e 1, D-70569 Stuttgart, Germany}
\affiliation{Max Planck Society Outstation at the Heinz Maier-Leibnitz Zentrum (MLZ), D-85748 Garching, Germany}

\author{V.~V.~Ustinov}
\affiliation{Institute of Metal Physics, 620180 Ekaterinburg, Russia}

\author{V.~L.~Aksenov}
\affiliation{Joint Institute for Nuclear Research, 141980 Dubna, Russia}

\author{B.~Keimer}
\affiliation{Max-Planck-Institut f\"ur Festk\"orperforschung, Heisenbergstra\ss e 1, D-70569 Stuttgart, Germany}
\date{\today}

\begin{abstract}
We have used spin-polarized neutron reflectometry to investigate the magnetization profile of superlattices composed of ferromagnetic Gd and superconducting Nb layers. We have observed a partial suppression of ferromagnetic (F) order of Gd layers in [Gd($d_F$)/Nb(25nm)]$_{12}$ superlattices below the superconducting (S) transition of the Nb layers. The amplitude of the suppression decreases with increasing $d_F$. By analyzing the neutron spin asymmetry we conclude that the observed effect has an electromagnetic origin - the proximity-coupled S layers screen out the external magnetic field and thus suppress the F response of the Gd layers inside the structure. Our investigation demonstrates the considerable influence of electromagnetic effects on the magnetic properties of S/F systems.
\end{abstract}


\maketitle

Artificial heterostructures with alternating superconducting (S) and ferromagnetic (F) layers are currently attracting  great attention due to a diverse set of proximity effects \cite{BuzdinRevModPhys,GolubovRMP,BergeretRMP,eschrig2011,SidorenkoLTP17}, including the Larkin-Ovchinnikov-Fulde-Ferrell phase, $\pi$-phase superconductivity and triplet pairing. These effects show how ferromagnetism influences the superconducting properties of the S/F heterostructures. Converse proximity effects in which superconductivity influences ferromagnetism have received less attention. Such magnetic proximity effects are expected in systems where the F and S transition temperatures, $T_F$ and $T_c$, are comparable, which is the case for heterostructures of cuprate high-$T_c$ superconductors and ferromagnetic manganates \cite{StahnPRB05,ChakhalianNatPhys06,HopplerNatMat09,SataphyPRL2012}, and for some bulk compounds \cite{aoki2001,pfleiderer2001,Mineev_2017}. However, because of the chemical and electronic complexity of these materials, simple model systems for magnetic proximity effects are highly desirable.

For most S/F heterostructures composed of elemental metals or alloys, $T_F$ greatly exceeds $T_c$. In such systems, one still expects significant magnetic proximity effects if the effective energy $E_{F} \sim T_F d_F/d_S$ becomes comparable to $E_S \sim T_c$, where $d_F$ ($d_S$) are the thicknesses of the F (S) layers \cite{BuzdinRevModPhys}. The first to indicate such a possibility were Anderson and Suhl \cite{AndersonSuhl}. They considered systems consisting of S and F phases and came to the conclusion that a homogeneous magnetic phase above $T_c$ may become inhomogeneous below $T_c$. Such a transition, which they called cryptoferromagnetism (CFM), would depress the effective exchange field of ferromagnetism, thus enabling the co-existence of superconductivity and magnetism. Later on the concept of CFM was further investigated in the theoretical work of Buzdin and Bulaevsky \cite{buzdin1988} and Bergeret et al. \cite{BergeretPhysRevB2008}. Recently Zhu et al. reported the observation of an increased coercivity below $T_c$ in GdN/Nb/GdN trilayers \cite{zhu2017superconducting}. The authors interpreted this increase as a superconductivity-driven antiferromagnetic (AF) alignment of the GdN layers.

The ability to control the magnetic state by superconductivity is attracting attention also in applied research on superconducting spintronics \cite{soloviev2017beyond,golubov2017superconductivity} including such new approaches  as neuromorphic computing \cite{schneider2018tutorial,soloviev2018adiabatic,schegolev2016adiabatic}. At the moment, most research efforts are focused on simple S/F structures such as bilayers and trilayers \cite{lenkBJN16,LenkPRB17,NevirkovetsPRA18}. However, both the superconducting and the magnetic  properties of more complex S/F systems, such as [S/F]$_n$ superlattices, may qualitatively differ from the properties of their S/F unit cells thus opening up perspectives for novel functionalities. An essential difference in behavior is expected when the thickness of the S and/or F layer  becomes comparable  with the coherence length of superconductivity in the S ($\xi_S$) or F ($\xi_F$) layers \cite{ProshinPRB2001,HaltermanPRB2004,Bakurskiy2015,nevirkovets2018,klenov2018periodic}. Preparation of such superlattices requires proper choice of materials with thin F and S layers, transparent S/F interfaces, and uniformity of the layer characteristics throughout the entire structure.

\begin{figure}[htb]
\centering
\includegraphics[width=\columnwidth]{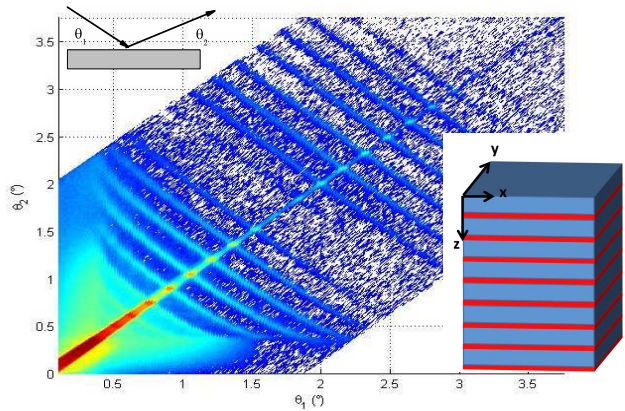}
\caption{
 X-ray scattering map from sample 2. The upper inset shows the scheme of the reflectometric experiment. The bottom inset shows a sketch of the structure where blue and red color indicate Nb and Gd layers.
 }
\label{Fig1}
\end{figure}

 Our recent study of Nb(25nm)/Gd($d_F$)/Nb(25nm) trilayers has shown that high quality structures with highly transparent S/F interfaces and rather high correlation length $\xi_F$ = 4nm can be grown \cite{KhaydukovPhysRevB18}. Moreover gadolinium is a weak ferromagnet with $T_{F} =$ 293K which in combination with Nb, the strongest elemental superconductor with $T_c = $9.3K, allows for preparation of S/F systems with $E_F \sim E_S$. In this work we report on a study of the magnetic and superconducting properties of superlattices of composition [Gd($d_F$)/Nb(25nm)]$_{12}$. The superlattices were deposited on 25x25mm$^2$ (1$\bar{1}$02)Al$_2$O$_3$ substrates  and covered by a Nb(5nm) capping layer. Later on we cut $\sim$5x5mm$^2$ pieces for magnetization and transport measurements (details of the sample preparation can be found in Ref. \cite{KhaydukovPhysRevB18}). The thickness of the Gd layers was chosen to be $d_F$ = 0.5$\xi_F$ (sample 1), 0.75$\xi_F$ (sample 2) and 1.25$\xi_F$ (sample 3). Fig. \ref{Fig1} shows an X-ray scattering map measured on sample 2 at the NREX neutron/X-ray reflectometer (details can be found in Ref. \cite{KhaydukovJAP15}).  The specular channel $\theta_2=\theta_1$ exhibits more than 15 Bragg peaks arising from diffraction on the superlattice with period $D=d_F+d_S$, demonstrating high repeatability of a Gd/Nb bilayer in the z-direction. In addition we have detected diffuse scattering for $\theta_2 \ne \theta_1$ in the form of tilted lines around the specular Bragg peaks. These Bragg-like sheets indicate a high statistical correlation of the in-plane roughness profiles of Nb/Gd interfaces in the periodic structure \cite{HolyBaumbach}.

\begin{figure*}[htb]
\centering
\includegraphics[width=2\columnwidth]{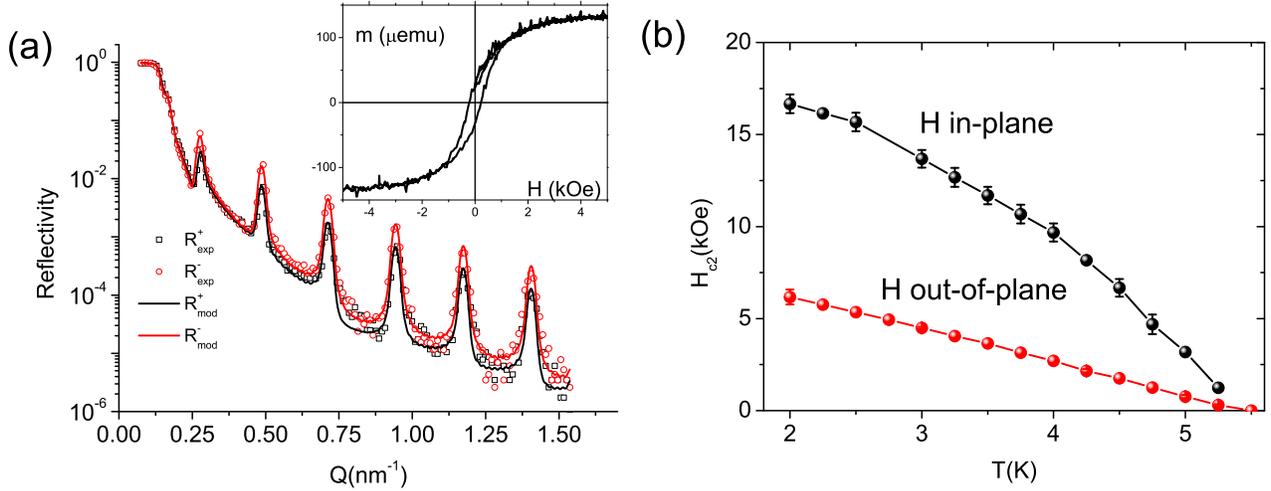}
\caption{
(a) Experimental (dots) neutron reflectivity curves measured on  sample 1 at $T$=7K and $H$ = 4.5kOe. Solid lines show model curves. Inset: Magnetic hysteresis loops measured at $T$ = 7K by a SQUID magnetometer. (b) Temperature dependencies of the upper critical field measured with external field applied parallel (black) and normal (red) to the surface.
}
\label{Fig2}
\end{figure*}

To study the magnetic properties of our superlattices we used  Polarized Neutron Reflectometry (PNR), which has been widely used for the study of S/F systems \cite{GoffJMMM2002,AksenovPhysicaB2005,StahnPRB05,ChakhalianNatPhys06,HopplerNatMat09,KhaydukovJSNM11,KhaydukovJetpLet13,SataphyPRL2012}. PNR allows measurements of the depth profile of the in-plane magnetization with nanometer depth resolution. Fig. \ref{Fig2}a shows typical  spin-up $R^+$ and spin-down $R^-$ reflectivity curves measured on sample 1 at $T$ = 7K (that is, in the normal state above $T_c$) and $H$ = 4.5kOe after cooling the sample in the same field. The neutron reflectivity exhibits six Bragg peaks positioned at $Q_n \approx 2 \pi n / D$. The difference of $R^+$ and $R^-$ clearly indicates the presence of an in-plane magnetic moment. Using the Born approximation one can show that the spin asymmetry at the $n$-th Bragg peak $S_n \equiv [R^+(Q_n) - R^-(Q_n)]/[R^+(Q_n) + R^-(Q_n)]$ is proportional to the magnetic contrast $M_{Gd}-M_{Nb}$ of a unit cell, where $M_{Gd,Nb}$ is the in-plane magnetization of a Gd and Nb layer, respectively. In the first approximation we may neglect the small magnetization of the Nb layers ($M_{Gd} \gg M_{Nb}$) and write $S_n \sim M_{Gd}$. We were able to reproduce the experimental curves with a model based on 12 identical pairs with $d_F$ = 1.7 nm and $d_S$ = 25.0 nm and magnetization of Gd of 4$\pi M_{Gd}$ = 2.5 kG. Within the measurement error of ~10\%, this value agrees with 2.7kG which can be calculated using the saturation magnetic moment $m_{sat} $=134 $\mu$emu measured by a Superconducting Quantum Interference Device (SQUID) magnetometer (inset in Fig. \ref{Fig2}a) at the same temperature and magnetic field.

Fig. \ref{Fig2}b shows the superconducting phase diagram of sample 1 measured by a standard four-point electrical resistivity technique. This phase diagram allowed us to determine $T_c$=5.5K and also the superconducting coherence length $\xi_S = \frac{2}{\pi} \sqrt{\frac{\Phi_0}{2\pi H_{C2 \perp} (0)}}$ = 11.6nm. The latter value is in agreement with the previously reported value for Nb films \cite{KhaydukovPhysRevB18}.

\begin{figure*}[htb]
\centering
\includegraphics[width=2\columnwidth]{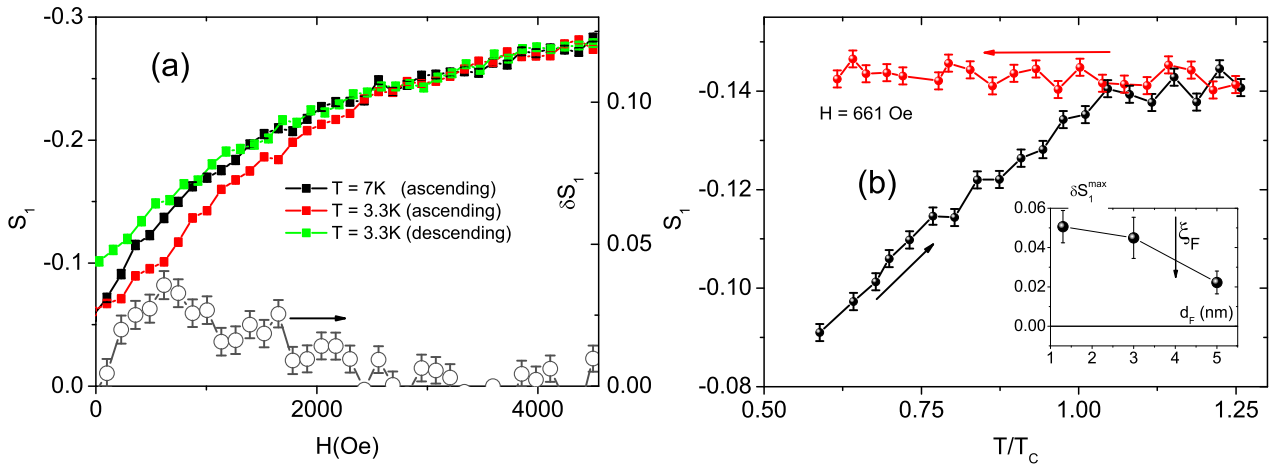}
\caption{
(a) Magnetic field dependent spin asymmetry at the first Bragg peak, $S_1$, measured above $T_c$ (black) and below $T_c$ (red and green). The black and red curves were measured in ascending magnetic field from zero to $H$=4.5kOe. The green curve was measured in descending field from saturation to zero. The sample was cooled in zero field below $T_c$ = 5.5K. Open symbols show the difference $\delta S_1$= $S_1$(7K)- $S_1$(3K). (b) Temperature dependence of $S_1$ measured in the magnetic field $H$ = 661Oe where the maximum of  $\delta S_1$ was observed. The black curve was measured by increasing the temperature after ZFC and after this the red curve was measured with decreasing temperature. The inset shows the $\delta S_1^{max}(d_F)$ dependence.
}
\label{Fig3}
\end{figure*}

Fig. \ref{Fig3} shows the field and temperature behavior of the spin asymmetry at the first Bragg peak $S_1$.  Above $T_c$ we used the following protocol. First the sample was magnetized for a short time in the maximum magnetic field $H_{max}$ = 4.5kOe. Then the field was released to zero and $S_1$(H) was measured for ascending magnetic field from 5Oe to $H_{max}$ (black curve in Fig. \ref{Fig3}a).  Then the field was released to zero and the sample was cooled down to T = 3.3K in zero magnetic field. After this we repeated $S_1$($H$) by first raising the magnetic field from 5Oe to $H_{max}$ (red curve in Fig. \ref{Fig3}a) and then lowering it to $H$ = 5Oe (green curve in Fig. \ref{Fig3}a). The $S_1(H)$ curve above $T_c$ repeats qualitatively the behavior of the upper branch of the macroscopic magnetic moment (inset in Fig. \ref{Fig2}a): the $S_1$($H$) curve grows from remanence to $H \sim$ 2kOe and then approaches saturation. The corresponding curve below $T_c$ is somewhat suppressed in the range of fields between remanence and saturation. The suppression is maximal around $H \approx $ 700 Oe. The descending curve in turn is different at small fields close to zero. In order to check whether this difference is related to the superconducting state we measured the temperature dependence $S_1$($T$) using the following protocol. Above $T_c$ the sample was magnetized to saturation for a short time and then the field was released to zero and the sample was cooled down to 3.3K in zero field. Then a field of $H$ = 661Oe was applied and $S_1$($T$) was measured by first heating the sample to $T$ = 7K (black curve in Fig. \ref{Fig3}b) and then cooling it back in the same field to $T$ = 3.3K. One can see that the amplitude of $S_1$ is indeed suppressed below $T_c$ if the sample is cooled in zero field. We have conducted similar measurements for the other two samples and observed that the magnitude of the suppression is inversely proportional to $d_F$ (see the inset in Fig. \ref{Fig3}a). For sample 3 with $d_F$=1.25$\xi_F$ the effect is small but non-vanishing.

We have thus observed a suppression of the spin asymmetry of the first Bragg peak below $T_c$ after zero-field cooling. The effect takes place in an intermediate range of magnetic fields between remanence and saturation. However, we did not observe any additional Bragg peaks below $T_c$ at these magnetic fields. Thus AF alignment or any other modification of the magnetic period can be excluded in our structure. Moreover in our previous study of Nb(25nm)/Gd($d_F$)/Nb(25nm) trilayers we did not observe any statistically significant change of the spin asymmetry below $T_c$ \cite{KhaydukovPhysRevB18}.  All these observations point to an electrodynamical origin of the effect.  Indeed for $d_F \sim \xi_F$ two adjacent S layers are expected to be coupled by the proximity effect. This means that the whole sample  is a superconductor with thickness $D_S = 12D \approx$ 300nm which is larger than the magnetic screening length $\lambda_{Nb}\sim$ 120nm  in niobium films \cite{ZhangPRB95,GubinPRB2005,KhaydukovJSNM2015}. Such a thick superconductor is able to expel a certain amount of external field. As a consequence, the central Gd layers feel less magnetic field than applied outside and hence their response is smaller. If the sample is cooled in a magnetic field, then magnetic flux is trapped around the Gd layers and the effect is smaller or not seen at all. This model also explains the existence of the effect in the intermediate range of magnetic fields where the  derivative $dM/dH \not= 0$. Moreover, the same mechanism explains why we didn't see the effect in Nb(25nm)/Gd/Nb(25nm) trilayers - the total thickness of the superconductor $D_S$ =50nm is not enough to expel a significant amount of magnetic flux.

\begin{figure*}[!htb]
\centering
\includegraphics[width=2\columnwidth]{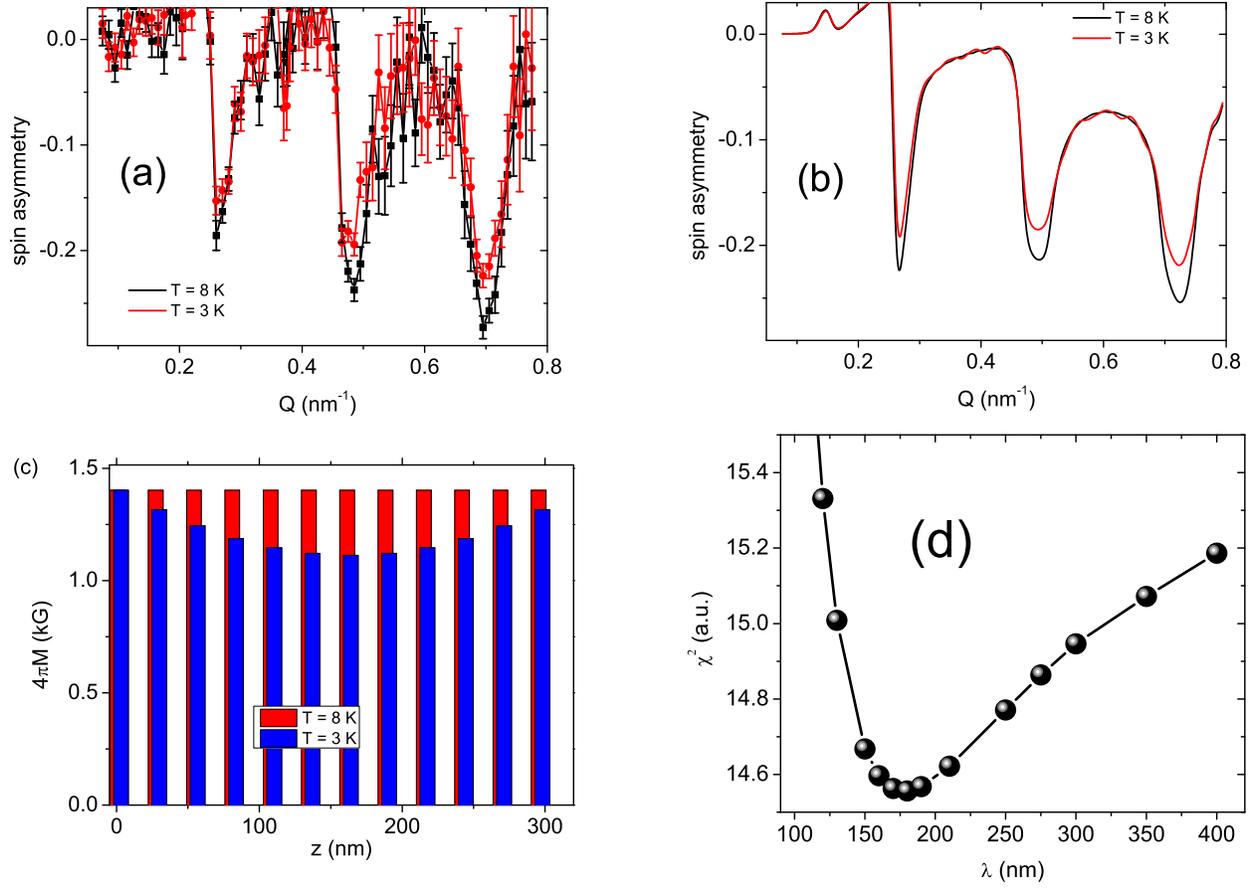}
\caption{
Experimental spin asymmetry measured on  sample 1 above and below $T_c$. (b) The best-fit model spin asymmetries for the magnetization depth profiles depicted in (c). (d) Dependence of the goodness of fit versus magnetic field penetration depth.
}
\label{Fig4}
\end{figure*}

In order to qualitatively describe the suppression of the spin asymmetry we have fitted the neutron data measured on sample 1 above and below $T_c$ in magnetic field $H$ = 0.8kOe (Fig. \ref{Fig4}a,b).  Above $T_c$ we used a model with twelve identical Gd layers. The best fit was obtained for $4\pi M_{Gd}$ = 1.4kG. To fit the data below $T_c$ we used the following procedure.  For the given value of $\lambda$ we calculated the value of the local magnetic field $H$($z$) at the position of every Gd layer using the well-known expression for the Meissner effect in a superconducting film of thickness $D_S$ and applied magnetic field $H_0$: $H(z) = H_0 ch(z/\lambda)/ch(D_S/2\lambda) $. The magnetic response of every Gd layer was then recalculated using this value of $H$($z$) under the assumption that all Gd layers follow the $M$($H$) dependence depicted in the inset to Fig. \ref{Fig2}a. Then model reflectivity curves for this magnetic configuration were calculated and compared to the experimental curves using the standard goodness-of-fit parameter $\chi^2$. Results of this treatment are shown in Fig. \ref{Fig4}, and the best agreement is obtained for $\lambda =$180 $\pm$ 10 nm (Fig. \ref{Fig4}d). This value is considerably larger than $\lambda \sim$ 100nm for a pure Nb film of similar thickness, which is typical for proximity-coupled multilayers  \cite{YusufJAP98,GalaktionovPRB03,HouzetPRB09}.

We have thus shown that thanks to the proximity effect the entire Gd/Nb superlattice behaves as a uniform thick (magnetic) superconductor. As a thick superconductor it is able to screen the applied magnetic field, thus suppressing the ferromagnetic response of the inner Gd layers. This effect is to some extent similar to the cryptoferromagnetism predicted in \cite{BergeretPhysRevB2008,buzdin1988,AndersonSuhl}. In analogy to CFM it leads to a transition from homogeneous magnetic order above $T_c$ to inhomogeneous order (along z in our case) below $T_c$, and hence to a suppression of the averaged magnetic moment. Similar to CFM, the effect takes place for a weakened ferromagnet ($d_F < \xi_F$) and strengthened superconductor ($D_S>\lambda $). Our investigation shows that electromagnetic effects may play a significant role in S/F systems and should be taken into account when considering proximity effects in S/F systems. Note that recent theoretical work came to a related conclusion \cite{mironovAPL2018}. Our results demonstrate the potential of elemental S/F multilayers as simple model systems for ferromagnetic superconductors.

The authors would like to thank G. Logvenov, G. Christiani, A. Melnikov and S.Mironov for fruitful discussions. This work is partially based on experiments performed at the NREX instrument operated by the Max Planck Society at the MLZ), Garching, Germany, and supported by DFG collaborative research center TRR 80. Research in Ekaterinburg was performed within the state assignment of Minobrnauki of Russia (theme "Spin" No. AAAA-A18-118020290104-2) and was partly supported by RFBR (Project No. 19-02-00674).

\bibliography{GdNb_Refs}

\end{document}